\documentclass[10pt,preprint]{aastex}

\slugcomment{To appear in PASP}

\shorttitle{SkyMapper Filters}
\shortauthors{Bessell et al.}

\begin{document}


\title{SkyMapper Filter Set: Design and Fabrication of Large Scale Optical Filters}


\author{Michael Bessell, Gabe Bloxham, Brian Schmidt, Stefan Keller, Patrick Tisserand \& Paul Francis}
\affil{RSAA, Mount Stromlo Observatory, The Australian National University, ACT 2611, Australia}
\email{bessell@mso.anu.edu.au}

\begin{abstract}
The SkyMapper Southern Sky Survey will be conducted from Siding Spring Observatory with u, v, g, r, i and z filters that comprise  glued glass combination filters of dimension 309x309x15 mm. In this paper we discuss the rationale for our bandpasses and physical characteristics of the filter set. The u, v, g and z filters are entirely glass filters which provide highly uniform band passes across the complete filter aperture. The i filter uses glass with a short-wave pass coating, and the r filter is a complete dielectric filter. We describe the process by which the filters were constructed, including the processes used to obtain uniform dielectric coatings and optimized narrow band anti-reflection coatings, as well as the technique of gluing the large glass pieces together after coating using UV transparent epoxy cement. The measured passbands including extinction and CCD QE are presented. 
\end{abstract}


\section{Introduction}
The success of the Sloan Digital Sky Survey (\url{http:www.sdss.org})\citep[eg.][] {Abaz09} and the immense amount of uniform and precise photometric data made available leads anyone intending to conduct a large scale survey to consider duplicating the SDSS passbands. However, as much of the SkyMapper \citep{Kell07} science case involves stellar photometry it is advantageous to design the filter set to provide  better discrimination of metallicity, effective temperature and effective gravity in stars than is provided by the SDSS filterset. There are older photometric systems which successfully achieve these goals \citep[see][]{Bess05}, namely the Stroemgren uvby system and the DDO system. With these insights we have moved the SDSS u and g bands apart and inserted a violet v band between them, but kept the redder passbands (r,i, and z) largely unchanged.

By shifting the SDSS u band further to the UV, we diminish the contribution of light redward of the Balmer Jump thereby emulating the Stroemgren u band. This bandpass is demonstrated to provide good temperature sensitivity in hot stars and good gravity sensitivity in A,F, and G stars. By shifting the blue edge of the g-band a little redward we lessen its metallicity sensitivity and make it behave more like Stroemgren b or Johnson V. Finally, in the gap between the redefined u and g bands, we introduce a v band similar to the DDO 38 band, which is very metallicity sensitive, especially at low metallicities. 

To facilitate the photometric precision and astrophysical fidelity of the SkyMapper photometric catalogs it is important that the passbands are consistent
across the filter aperture and that their transmission does not change with time. This is especially true in the u and v bands, whose power to accurately discriminate stellar parameters depends on precision photometry. While interference filters can be made in large sizes, they remain very expensive, with uniformity of the passband across the aperture difficult to achieve.  To overcome these drawbacks, interference filter manufacturers suggest making a mosaic of smaller filters rather than a single large filter, something we did not wish to do given our monolithic focal plane assembly. There are also problems with wavelength stability of interference filters with time and environment. For all these reasons,  we investigated  and pursued fabricating large colored glass filters that we have successfully used in the past for long-term photometric programs with single CCD systems. 

In section 2 we will discuss color glass filters and describe choosing the glasses and sourcing the blanks. In section 3 we will describe grinding and polishing the glass to the required thickness and surface finish, applying the SWP and anti-reflection (V-AR) coatings and gluing the glass assemblies.  In section 4 we will present the resultant passbands and in the final section provide a summary.

\section{Colored glass filters}
The palette of colored glass filters is quite extensive although the range of glass types has been reduced since the review paper on colored filter glasses by \citet{Dobr77}. 
Much of the discussion in that paper is still relevant, but whereas in 1977 there were 13 manufacturers of colored filter glass in 6 countries, today we have knowledge 
of only 5 companies still producing filters, although companies such as Schott have facilities in several different countries. As well as a reduction in the range of glass type, the forms of the glass production (size of sheets and thicknesses) and method of production has also changed. The Schott filter catalog \footnotemark
\footnotetext[1]{\url{http://www.schott.com/advanced\_optics/english/download /schott\_optical\_glass\_filters\_2009\_en.pdf}}
provides a good summary of the chemical/physical attributes of the glasses that generates the variety of colors. 
Briefly, colored glasses are produced in one of two ways, either by ionic 
coloration or by absorption and scattering from a suspension of
colloidal particles that are produced in the glass and controlled in size 
by heat treatment after an essentially colorless glass is made. 

The ionic glasses are made by dissolving particular salts (such 
as cobalt or nickel oxide) in glass. The ionic coloration of the UG (violet) 
and BG (blue) series of glasses produces a spectral transmittance curve 
resembling a bell-curve of half-width between 100 and 200 nm but most of 
these glasses also transmit red light beyond 700 nm as well, thus the 
violet glasses appear purple to the eye.    

The second type of colored filter glass are a series of sharp edge 
(or short wave cut-off) filters, the WG, GG (sulphur and cadmium sulfide), 
OG (cadmium selenide) and 
RG (gold) series which absorb light blueward of a quite sharply defined
wavelength. The filters are made with their short wave cutoffs ranging from
400 to 800 nm in steps of 20 nm. To the eye, the WG glasses are essentially 
colorless, the GG series go from colorless through light yellow to dark
yellow, the OG are orange-red and the RG series are rose to ruby in color. Selecting a thicker piece of glass moves the shortwave cutoff redward. 

Schott also make special (nearly) colorless glasses (BG39/40, S8612 and 
the KG series) that can be used as long wave cutoff filters; however, 
these glasses have long wave cutoffs that fall much more slowly 
(over about 250 nm) than the short wave cutoffs rise (over about 50 nm) thus producing an asymmetrical 
spectral transmittance curve when used together to define a bandpass. 
An important use for the BG39/40 and S8612 glasses is also as  a "red-leak" 
blocker, absorbing light beyond 700 nm where most of the blue and 
violet glasses transmit. The better UV transmission of Schott S8612 compared to BG39/40 
makes it the preferred choice to block the red leak of the UG glasses. 

There are two main filter manufacturers apart from Schott, Hoya (\url{http://www.hoyaoptics.com/color\_filter/index.htm})
and Nantong Yinxing Optical (\url{http://www.ygofg.com/Products.htm}) both whose product lines are similar
to that of Schott.   There are also at least two manufacturers in the former USSR (Russia and Ukraine). Yinxing Optica provides a useful cross reference of Chinese glasses and other manufacturers , including  glasses from the former USSR (\url{http://www.ygofg.com/Cross.htm}).

\subsection{Sources of large filter glass}
Two important requirements for SkyMapper filters are their large size, 309x309 mm, and large thickness, 15 mm -  or more precisely, the specific filter glass thickness (depending on its refractive index) that equates to 15 mm of BK7 glass. Using a thicker filter mitigates ghost intensities and makes fabrication and assembly easier. Unfortunately, Schott, Hoya, and Yinxing production lines currently only fabricate filter glass in strips up 165mm wide, with a maximum 3mm finished thickness. \footnotemark
\footnotetext[2]{Schott is considering making filters up to 350x350x8mm again \citep{Doeh06}. GG, OG and RG type glasses can still be made by special melts in large sheets, but still with a limited thickness.} 
So we looked to companies in Russia to see whether filter glass was still made in large slabs. Historically, in the former USSR different colored filter glasses were made in several different plants but they worked to identical standard formulations and met standard catalogue transmission values.  A listing of the GOST 9411-81 standard catalogue of USSR filter glasses is available from \url{http://www.mso.anu.edu.au/$\sim$bessell/FTP/Filters}. A particular added advantage of these GOST filter glasses was that their filter range included glasses with a range of concentrations of the colorant enabling thicker glasses to be specified for the same color as a standard thin Schott filter.

We initially acquired filter glass from a Russian company MacroOptica who sourced most of their glass from the companies Krasnyi Gigant (since closed) and LZOS (\url{http://lzos.ru/en/}).   Since 2008, we have acquired red (RC10), clear (C3C21 $\sim BG40$) and UV (UVC2) glasses from Potapenko Glass \& Filters (PG\&F, TM) of the Ukrainian Optical Glass Factory who consistently delivered glasses of high quality. Links to datasheets of their GOST named filters and a list of equivalences to Schott and Hoya filters is available at \url{http://www.opticalglass.com.ua/en/catalog/filtersheet.htm}
Stocks of large filter slabs are not usually held but are made to order. In the case of colloidal colored glasses, special melts require a long
lead time because the processing required to  color-strike and anneal large thick slabs of glass uniformly is time consuming. The LZOS company also offered to make large filters in the yellow, orange and red glass type ranges. The range of the filters from PG\&F and LZOS is also given on  \url{http://www.mso.anu.edu.au/$\sim$bessell/FTP/Filters}.

Based on transmission values in the GOST 9411-81 catalog we designed glass mixes and thicknesses for the SkyMapper bands after verifying the catalog transmission values against 50 mm samples of the glasses that we ordered. Table 1 lists the filter recipes that we have used. The u band filter was originally 8mm y$\Phi$C2 + 6.3mm BC4 but we discovered  an optical flaw in the corner of the y$\Phi$C2 glass after the filter was assembled and tested on the sky. To expedite the filter fabrication, we then ordered some UVC2 from PG\&F and used a piece of Schott B270 that we had on hand to bring the thickness up to 15mm. Its thermal properties are similar to that of UVC2 and its cut-on edge was suitable. The resultant u passband differed slightly from the original formulation with a small red leak (intensity 0.7\%, 250$\AA$ wide centered at 7170$\AA$), but with a higher overall transmission of 60\%. The r filter was originally designed to use a piece of OC12 ($\sim$ GG550) from MacroOptics for the short wavelength cut-off but on delivery, the non-uniformity (color variation across the aperture) was clearly visible by eye. Due to time and financial constraints, as well as a lower requirement for uniformity in the r band, we decided to use dielectric coatings on BK7 glass for both sides of the passband. The resulting filter has very high throughput, but is less uniform than the glass filters. 

\begin{deluxetable}{cl}
\tabletypesize{\scriptsize}
\tablecaption{Filter glasses used in SkyMapper}
\tablewidth{0pt}
\tablehead{\colhead{Band} & \colhead {Component glasses} }
\startdata
u& 6.2mm UVC2 + 8.3mm B270 \\
v& 3mm y$\Phi$C1 + 6.5mmC3C23 + 5 BC7 \\
g& 4mm GG420 + 5mm C3C21 + 5.7mm BC4 \\
r& 15mm BK7 + LWP + SWP \\
i& 6mm KC19 + 8.5mm B270 + SWP \\
z& 4.5mm RG850 + 10.1mm B270  \\
\enddata
\tablecomments{B270 and BK7 are clear Schott optical glasses used to build up the thickness.
SWP and LWP indicates Short Wave Passband and Long Wave Passband coatings that were used to define the red or blue sides of the band.
The z filter was specially made by Schott.}

\end{deluxetable}

\section{Filter manufacture}
\subsection{Packing and shipping}
Glass is fragile and needs special packing, an obvious, but important point. When ordering glass, it is important that the shipping container be specified as part of the contract. We recommend that the smaller box containing the glass is supported by compressible material within another container. Our preferred method is to pack in bubble wrap inside a corrugated cardboard box that is itself packed in foam in a model 1610 Pelican case (\url{http://www.pelican.com}). 

\subsection{Cutting and Polishing}
To ensure that the filters do not impinge on the delivered image quality and uniformity of the telescope, we specified standard optical quality (ST) as regards homogeneity, wavelength tolerance, stress birefrigence, visible stria and visible bubbles. We also specified tolerances for the physical dimensions of the glass and for the final polish. The mechanical tolerances on the edge dimension were  $\pm$ 0.5 mm; on the thicknesses $\pm$0.08 mm and the wedge angle was to be less than 0.0074$^{\circ}$/0.04 mm. Optical tolerances were specified to be less than 0.5 waves peak-to-valley transmitted over any single beam footprint of 40 mm diameter within the optical clear aperture; the overall power of each surface to be less than 9 waves at 633 nm; and the surface irregularity to be less than 5 waves at 633nm with a surface finish of 40-20 SD (scratch and dig).  

Our first lot of glasses polished by MacroOptica did not achieve the polishing specifications. To rectify this problem we had the glasses  repolished by Bond Optics (New Hampshire) (\url{http://bondoptics.com}). Subsequent glasses from PG\&F were ordered in a rough ground state. To assess the optical transmission quality before polishing, we oiled each sheet of ground filter glass and placed it between sheets of plate glass.  Selected pieces were sent to Bond Optics for grinding to the required thickness, and then polishing. Bond  Optics were able to achieve our required specifications.

\subsection{Coating the filters}
Rather than use the shallow red cutoffs of the available glasses, we elected to define the long wavelength edges of the r and i bands by means of a SWP coating (due to the delivered piece of OC12 glass not meeting our specifications we also elected to define the short wavelength edge of the r filter with a LWP coating). After consultation with a number of manufacturers, we approached Reed Schmell of ATOC Optical Sputtering Technologies (ATOC/OST) (New Mexico) (\url{http://www.opticalsputtering.com}) who was able to offer the uniformity runoff across the aperture that we required using Modulated Reactive DC Magnetron Sputtering. OST designed and applied the required coatings. They achieved high on-band transmission, and although quite uniform, the filters had somewhat larger fluctuations than anticipated due to equipment failure during the actual deposition run. OST is also making a narrow-band H$\alpha$ filter (6563$\pm$40\AA) and a MgH filter (5142$\pm$80\AA) (to distinguish K giants from K dwarfs) on Ukrainian colored glass substrates that will also be used on SkyMapper. 

A simple two-layer "V-band" (shape of the coating passband) antireflection coating was applied to all the color glass filters by Optical Coating Associates (\url{http:www.opticalcoating.com.au}). The coating process involves heating the substrate to 70-75$^{\circ}$C to achieve a good base vacuum then using ion assisted deposition to apply the 2 layer VAR-coat over a 5-10 min period. The coating process introduces additional heat and the chamber registered more than 100$^{\circ}$ when the coating guns were switched off. In our first attempt, we  glued the component glasses together before coating.  However, the first filter coated, the g filter, partially delaminated due to the heat absorbed by the middle glass (C3C21$\sim$ BG40) which absorbs light at infrared wavelengths. Subsequently, we decided to make new filters, applying the VAR coating to the individual glass pieces  and gluing the filter assembly together as the final step.
\subsection{Cementing the filters}
Schott constructed the SkyMapper z filter using UV setting cement NOA 61. We could not use this technique on our other filters as UV setting cements are incompatible  with some colored glass types and also require very uniform large UV light cabinets for processing.  The cement also shrinks causing the glass to bend which often requires repolishing to recover the desired flatness. We have had much experience  successfully gluing smaller filters with UV transmitting epoxy cement, with the epoxy joints proving to be very stable over time.  
 
We use EPO-TEK 301-2FL (\url{http://www.epotek.com/sscdocs/datasheets/301-2FL.PDF}) that has excellent UV transmission properties.   It is a slow setting (3 days) optical epoxy that hardens with near zero shrinkage that puts little strain on the glasses and is easy to work with in a small laboratory environment.. This glue has a short shelf life and was ordered as required. As the glue ages, small translucent flecks of material form that do not redissolve on warming. 

The technique for successful gluing is straightforward but time consuming. We elaborate the procedure here. New dust-free vinyl gloves should be used whenever the filters are handled. The laboratory work area needs to be kept clear of dust and it is important that the air is filtered and the temperature kept about 26$^{\circ}$C. Whenever a new glue run is made, two pieces of plate glass the same diameter as the filters are glued first as a trial. This enables the correct amount of glue to be better estimated and the uniformity and behaviour of the glue layer to be closely monitored. The edges of the filters to be glued need to restrained in a rig to ensure that they stay aligned during the gluing process (Fig 1). However, access to the edges needs to be available so any glue that leaks out can be easily wiped off.  

At the start of the procedure, 5 pieces of mylar shim of thickness 0.1 mm are placed on top of each other at each corner of the filter. A precise amount of mixed epoxy is weighed and poured in a Z pattern, about 200 mm across, near the centre of the lower glass sheet. (The Z pattern is important at it prevents air bubbles being trapped in the spreading epoxy). The second piece of glass is then lowered with a controlled wedge action on top of the first. This is allowed to stand under its own weight forcing the glue (and any bubbles) slowly across the whole area. After 30 mins, the first shim is slid out  of one corner slightly lowering the glass on that corner. The other first level shims are then removed from the other corners in turn over the next hour, gradually allowing the glue to slowly flow/float between the plates out to the edge. The next shim layer is then removed in the same manner as the first layer, with the sequence of removals of each layer continued for 2-3 hours until the final layer of shims remain - one in each corner with the glue fully filling the glass all the way to the edge gap over the entirety of the surface (Fig 2). As the glue slowly thickens (the consistency of thick honey), but is not tacky, the final four shims are removed together in diagonal pairs. Glue continues to slowly leak out for a while, but an even layer of glue - about 30-40$\mu$ thick - remains between the two glasses when finally cured. All glue that leaks from the edges throughout the shim removal process is carefully wiped off making sure that none touches the surface of the filter.  After a 72 hour curing period the filter is ready for use.  This process produces a filter which is extremely stable and can withstand the test of time.

\section{Filter measurements for optical quality and bandpass uniformity}
\subsection{Optical quality tests}
The completed filters were held vertically and the optical quality assessed qualitatively  using a Ronchi screen at the center of curvature of a spherical mirror (radius 3000 mm \& 520 mm diameter). Each filter or filter glass components that had been polished plano, were inserted in the beam path adjacent to the spherical mirror and any local area imperfections in the color filter glass or distortions of the wavefront (double pass in transmission) were then easily identified by visual inspection of the Ronchi pattern. This is a very  simple and sensitive test.  Fig 3 shows the fringe pattern through a good and bad section of a rejected filter. A camera was used for the i and z filters but the u filter could not be tested this way and was instead assembled as spaced components in a filter cell and tested on the sky in SkyMapper.  Additional quantitative optical tests were performed with a collimator / camera arrangement with a 30 mm diameter footprint on the filter glass. Here a perfect object point source was imaged through a specific footprint area of the filter, previously identified in the Ronchi test above, and the image PSF was analysed, and the aberration calculated from that data.    

\subsection{Wavelength transmission uniformity}
The wavelength transmission of the filters was measured with an Ocean Optics HR4000 mini fibre optic Spectrometer, an Optic DT mini 2GS light source, a pair of 600 micron silica fibres and a pair of 74-UV collimating lenses. The useful range of the Spectrometer is from 250 nm to 1000 nm and our resolution was about 1nm. The filter was held vertically on an X-Y slide placed between the transmit and receive fibres and the X-Y slide moved to enable the probe beam ($\sim$ 4 mm diameter) to measure the wavelength transmission at 25 points across the aperture of the filter to measure the uniformity. The transmissions measured at 13 points over the aperture are shown superimposed in Fig 4. All filters exhibit good uniformity, especially the all-colored-glass filters that show almost no variation. The dielectric coating on the i filter also showed a small variation top to bottom across the aperture.

\subsection{Wavelength transmission variation with angle of incidence}
The filters in SkyMapper are used in the convergent beam and the extreme cone angles of the beams are different for different parts of the field \citep[eg.][]{Elli76, Park98} and the large secondary obscures the low angle incident rays. To assess the resultant wavelength shifts for the SkyMapper filters we measured the wavelength transmission variation with angle of incidence (AOI) using the above measuring setup, but with the filter and an identical clear (compensation plate) inserted in the measurement space as shown in Fig 5. The AOI was varied by changing the angle of the "V" arrangement, of filter and compensating plate. Spectral data was collected for a range of AOI out to approximately 12 degrees. A function describing the wavelength shift with AOI was then derived at the sample points across the aperture and this data was used to reconstruct the shift for the telescope cone beams in angle steps.

The dielectric coatings showed larger non-uniformities with viewing angle compared to the colored glasses. Fig 6 shows, for example,  a comparison of the variation of transmission with angle of incidence for the GG420 cuton of the g filter and the SWP cutoff of the r filter. We computed for the i filter, that the red edge of the i band for the axial cone shifts 13$\AA$ to the blue compared to the measured normal incidence ray. The shift is a further 3$\AA$ for the edge and 6$\AA$ for the corner of the field. The blue edge of the i band, being colored glass defined, scarcely moves. For the r filter, the blue and red edges of the axial cone shift blueward by 12$\AA$  and 6.5$\AA$, respectively.
To the first order, these variations in the wavelength transmissions of the r and i filters will be accounted for by the flat field, although in principle, systematic differences can arise for objects with large spectral changes near the edges of the passbands. As noted earlier, this would be of much more concern for the u, v and g bands, whose power to accurately discriminate stellar parameters depends on precision photometry. However, these three colored glass filters show exceptional angular and spatial uniformity. 

 \section{The SkyMapper Passbands}
 The mean of the filter traces were used as the filter passband. For the r and I bands, we used the computed axial cone values. The primed values in Table 2 are the measured transmission values of the filters. Fig 7 shows and Table 2 also lists the normalized passbands including the mean QE of the E2V CCD44-82 devices and one airmass of atmospheric extinction for Siding Spring Observatory (including mean telluric H$_{2}$O and O$_{2}$ bands). The CCD response shortward of 4000$\AA$ are more poorly determined than at longer wavelengths, so it is likely that the calculated SkyMapper passbands in Table 2 will need slight adjustment after on-sky observations are made of our primary spectrophotometric standard stars. As mentioned above, the u filter has a small red leak that is also given in Table 2. 
 
\subsection{Stability of passbands}
The transmission passbands of the ionic glasses in the u and v filters are stable with temperature but their surfaces have low chemical oxidation properties. However, as their surfaces are protected by the V-AR coating or by the epoxy cement, this feature should not be an issue. The wavelength transmission of the other color filter glasses do change with temperature but the shifts are small for the expected summer-winter variations. It is important to note that some of the colloidally coloured glasses (e.g. GG, OG) can have their bandpasses permanently altered by application of moderate amounts of heat, and such treatment (which occurs, for example, in attaching filters with molten pitch for polishing or in some methods of AR coatings) needs to trialled.
The V-AR coatings are hard but can be damaged by moisture. Our filters are stored in dry air, and during an exposure dry air is continually blown through the filter and corrector assembly. 
The LWP and SWP coatings on the r and i filters are  dense, low-absorption, low-scatter, high-energy laser optical coatings deposited by Modulated Reactive DC Magnetron Sputtering (MRDCMS) and exhibit much improved environmental stability and durability compared to more porous coatings produced by conventional multi-layer evaporative coating processes. We anticipate excellent wavelength stability from these coatings  but will monitor any transmission changes over time via SkyMapper calibration procedures.

\section{Summary}
Obtaining large sized interference filters which are both uniform and high in their transmission in bands spanning the ultra-violet to the far-red remain challenging and very expensive. Using a combination of colored glass and dielectric coatings, we have produced a highly uniform set of filters for the SkyMapper telescope for moderate cost. 
Reliable sourcing of colored glass was the principal difficulty we encountered, but the Ukrainian based Potapenko Glass \& Filters (PG\&F) company appears to be able to consistently supply the requisite glass in a timely manner. In addition, Schott are considering making large dimension filters again. Since we believe we are the first group to attempt to make large format glass filters, we have shared our experiences to make the process easier for others in the future. 
 \acknowledgments
 
 We wish to thank Thorsten Doehring of Schott AG (Germany) for specially making  a large sheet of the unique Schott glass RG850 and constructing the z filter for SkyMapper; and Ross Zhelem who very ably assisted with the original filter designs and verified the transmissions of the Russian glasses.  We also wish to thank the late Al Collins (Collins \&Assoc), Serge Potapenko (PG\&F), Bill Doherty (Bond Optics), Reed Schmell and Andres Rael (ATOC/OST) and David Baker (Optical Coating Associates)  for the extra care and interest that they took in working with us to make the excellent SkyMapper filters.

\begin{figure}
\plotone{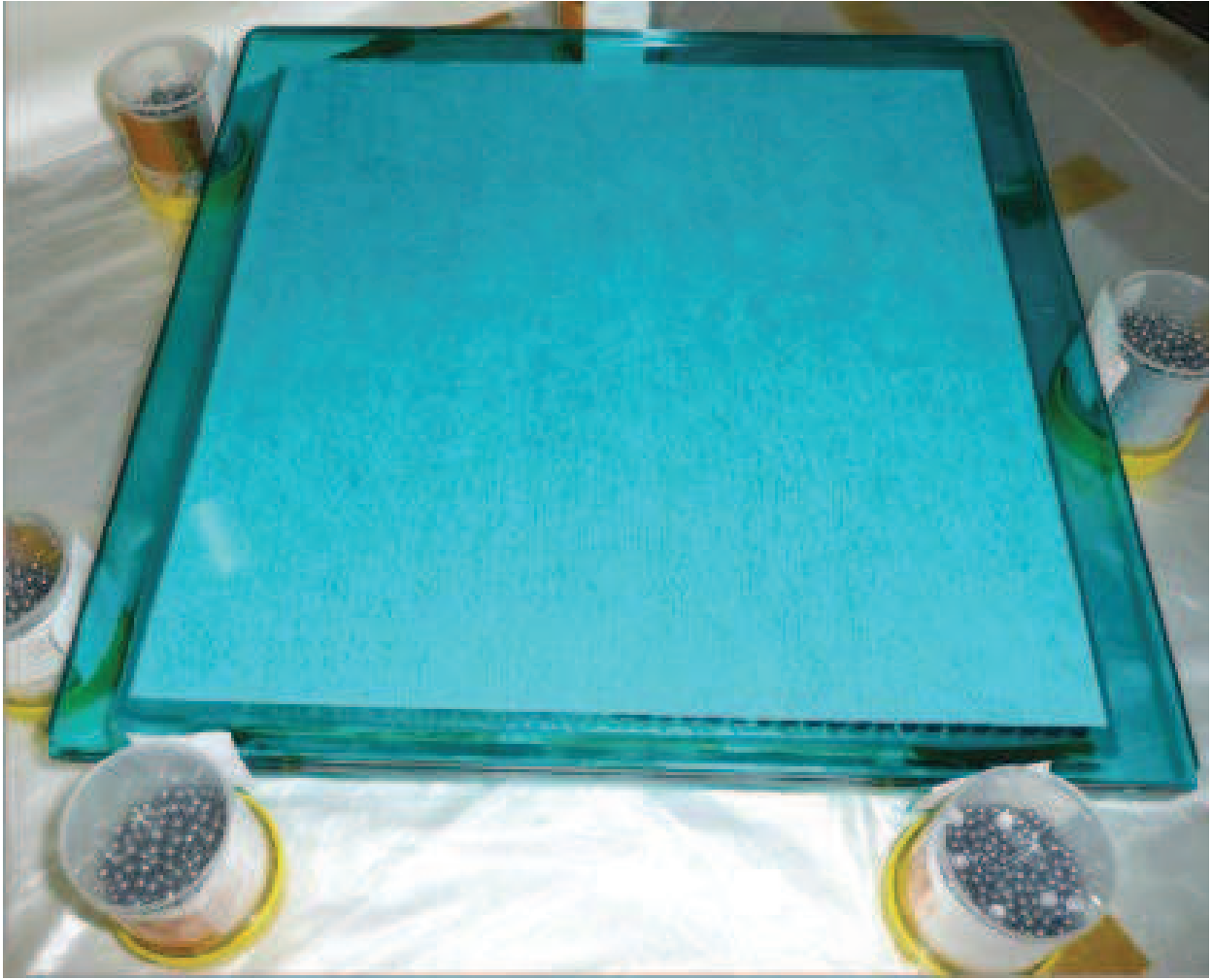}
\caption{Gluing the g filter}
\end{figure}

\begin{figure}
\plotone{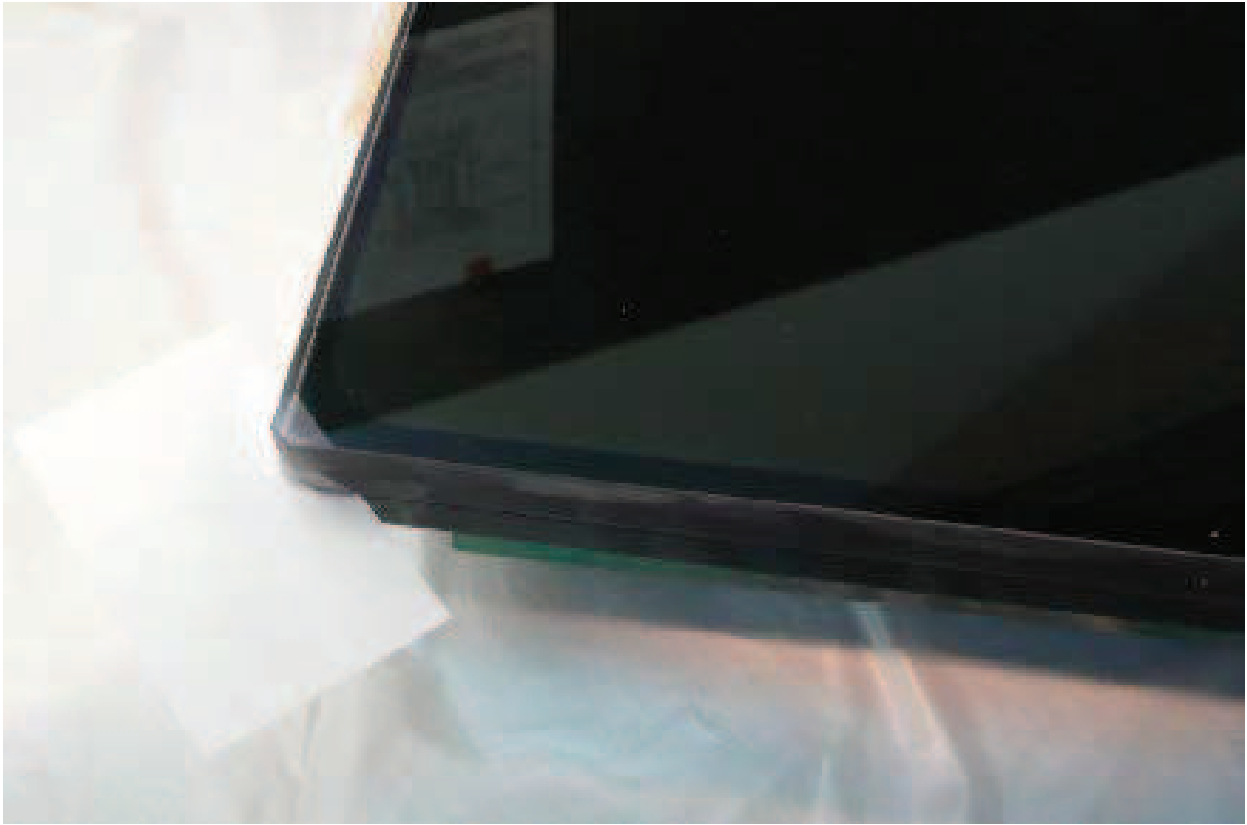}
\caption{The last mylar shim in place while gluing the u filter}
\end{figure}

\begin{figure}
\plotone{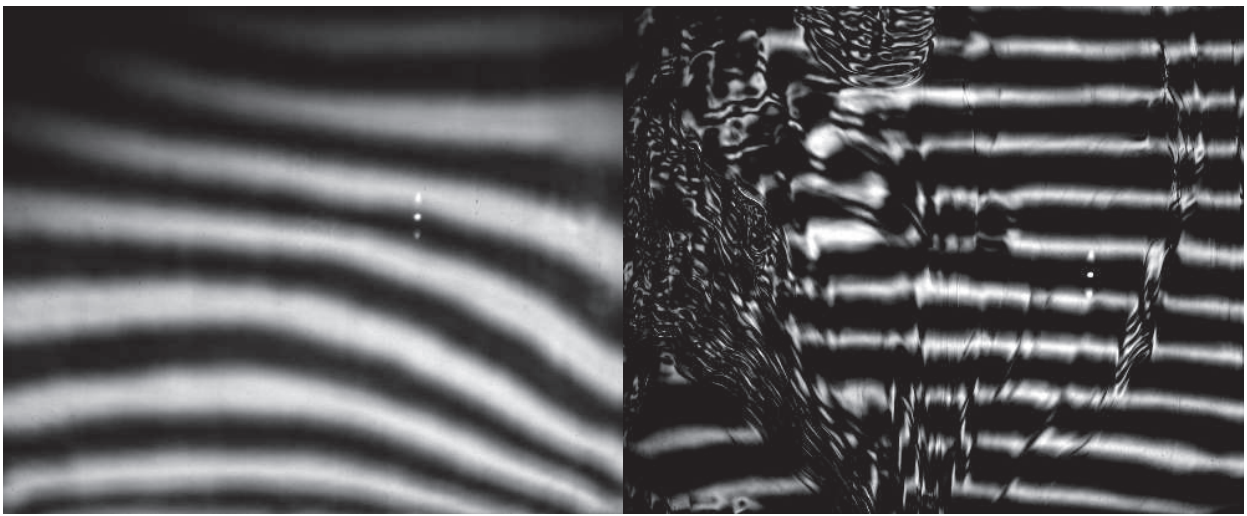}
\caption{Ronchi fringes through good and bad sections of a rejected filter.}
\end{figure}

\begin{figure}
\plotone{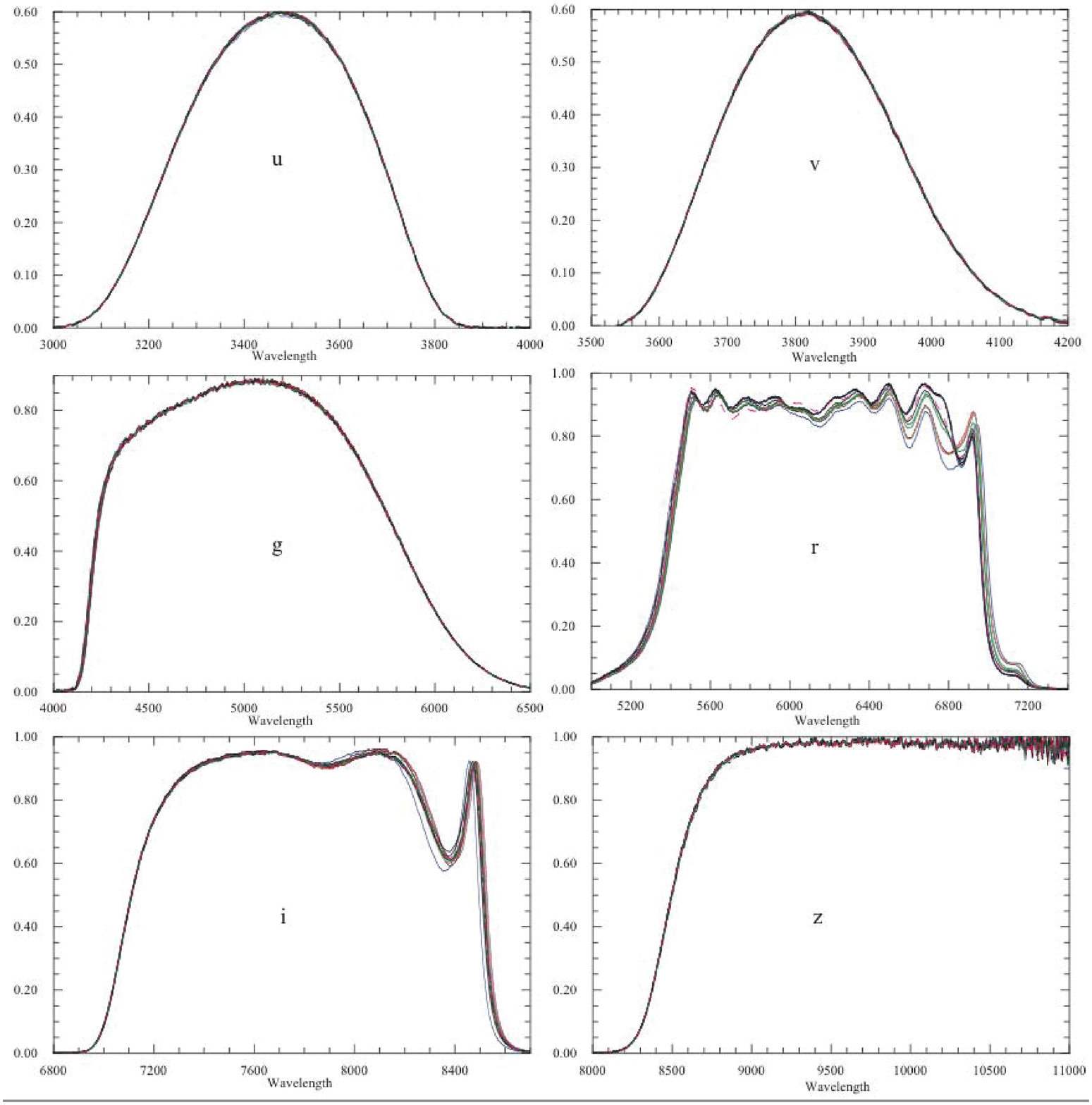}
\caption{The superimposed measured transmissions at 13 points over the aperture.}
\end{figure}

\begin{figure}
\plotone{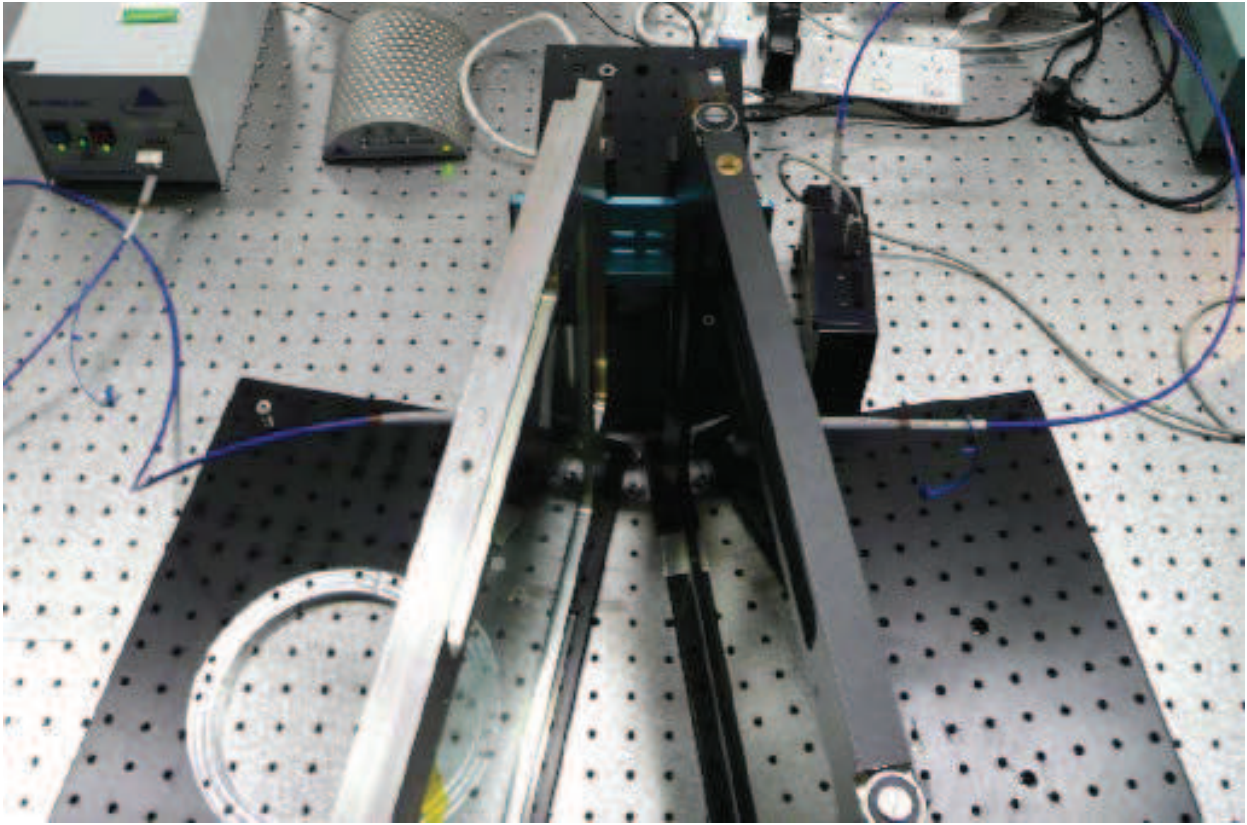}
\caption{Optical setup to measure transmission variation with incident angle.}
\end{figure}

\begin{figure}
\plotone{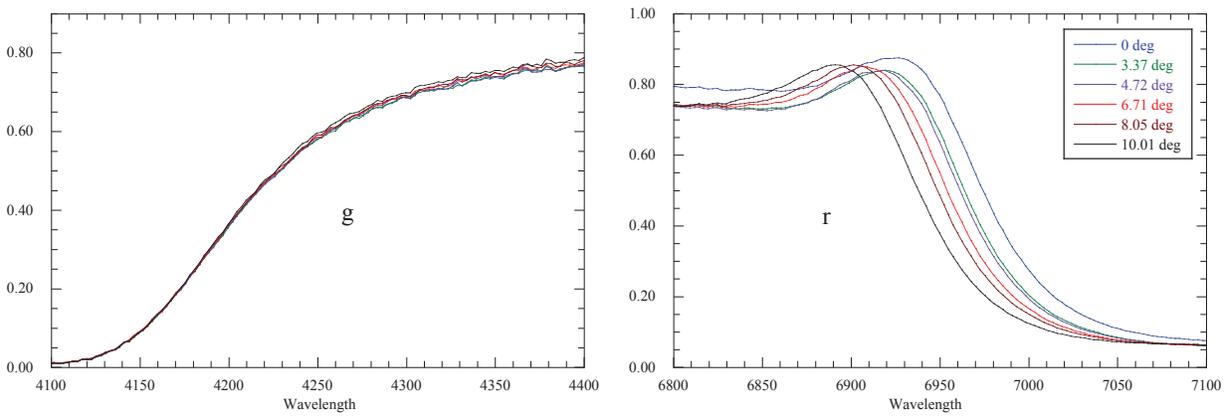}
\caption{The superimposed measured transmissions at six angles of incidence between 0 and 10 degrees of the g and r filters.}
\end{figure}

\begin{figure}
\plotone{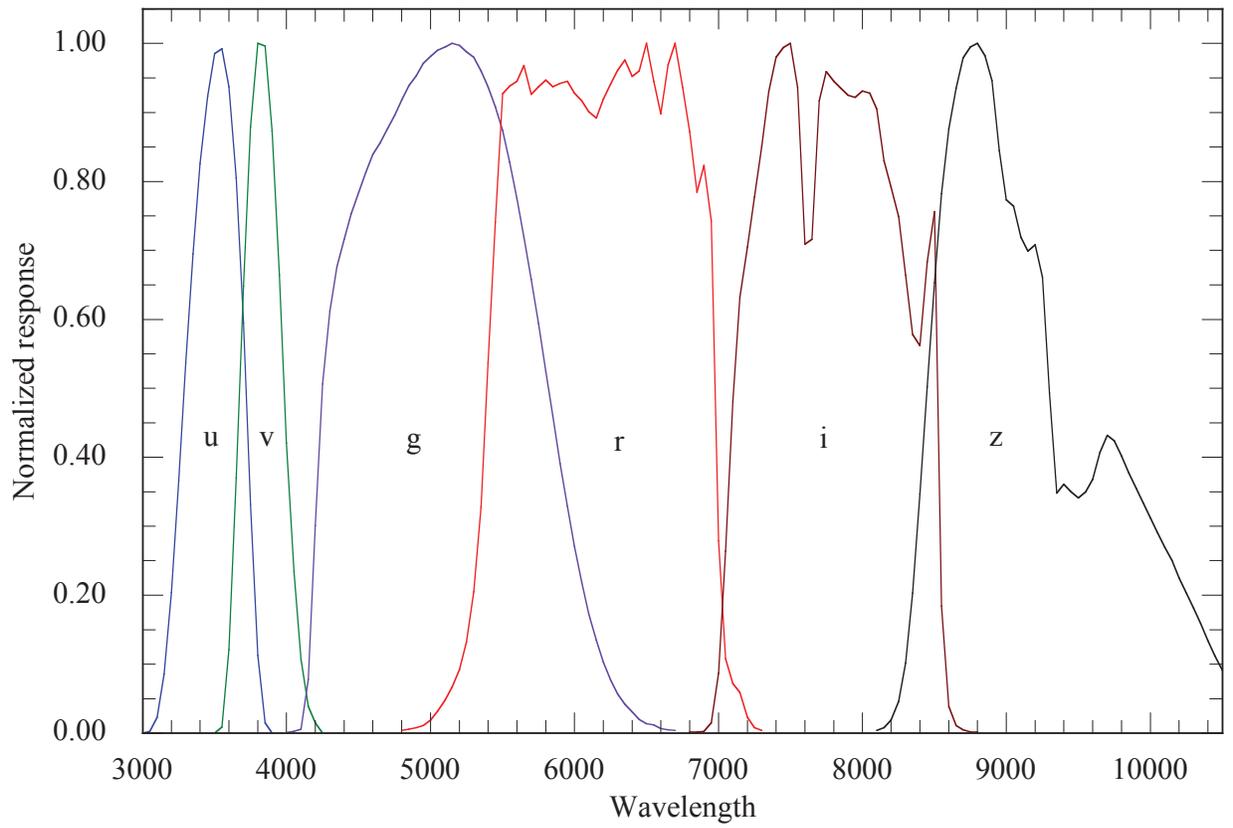}
\caption{Normalized SkyMapper response functions}
\end{figure}

\begin{deluxetable}{rrrrrrrrrrrrrrrrrr}
\tabletypesize{\tiny}
\tablecaption{SkyMapper normalized response functions and measured filter transmissions}
\tablewidth{0pt}
\tablehead{\colhead{Wave} & \colhead {u}& \colhead {u'} & \colhead{Wave} & \colhead {v} & \colhead {v'} & \colhead{Wave} & \colhead {g} & \colhead {g'} & \colhead{Wave} & \colhead {r} & \colhead {r'} & \colhead{Wave} & \colhead {i} & \colhead {i'} & \colhead{Wave} & \colhead {z}  & \colhead {z'}}
\startdata
3000 &0.000 &0.000 &3500 &0.000 &0.000 &4000 &0.000 &0.001 &4800 &0.000 &0.004 &6800 &0.000 &0.002 &8100 &0.000 &0.003 \\
3050 &0.003 &0.010 &3550 &0.009 &0.007 &4050 &0.003 &0.004 &4850 &0.002 &0.006 &6850 &0.001 &0.002 &8150 &0.008 &0.006 \\
3100 &0.023 &0.043 &3600 &0.121 &0.085 &4100 &0.006 &0.007 &4900 &0.008 &0.008 &6900 &0.003 &0.003 &8200 &0.019 &0.015 \\
3150 &0.086 &0.117 &3650 &0.370 &0.247 &4150 &0.078 &0.087 &4950 &0.012 &0.011 &6950 &0.015 &0.014 &8250 &0.046 &0.035 \\
3200 &0.204 &0.222 &3700 &0.648 &0.413 &4200 &0.301 &0.326 &5000 &0.021 &0.019 &7000 &0.087 &0.080 &8300 &0.102 &0.078 \\
3250 &0.366 &0.337 &3750 &0.879 &0.537 &4250 &0.506 &0.537 &5050 &0.034 &0.032 &7050 &0.262 &0.242 &8350 &0.203 &0.153 \\
3300 &0.537 &0.438 &3800 &1.000 &0.590 &4300 &0.612 &0.637 &5100 &0.051 &0.048 &7100 &0.480 &0.444 &8400 &0.347 &0.261 \\
3350 &0.695 &0.517 &3850 &0.996 &0.568 &4350 &0.676 &0.692 &5150 &0.070 &0.066 &7150 &0.633 &0.616 &8450 &0.502 &0.383 \\
3400 &0.826 &0.567 &3900 &0.873 &0.483 &4400 &0.716 &0.722 &5200 &0.097 &0.091 &7200 &0.707 &0.731 &8500 &0.653 &0.508 \\
3450 &0.923 &0.591 &3950 &0.664 &0.357 &4450 &0.753 &0.748 &5250 &0.141 &0.130 &7250 &0.777 &0.807 &8550 &0.782 &0.620 \\
3500 &0.985 &0.593 &4000 &0.421 &0.220 &4500 &0.784 &0.768 &5300 &0.218 &0.201 &7300 &0.851 &0.857 &8600 &0.876 &0.710 \\
3550 &0.992 &0.565 &4050 &0.234 &0.119 &4550 &0.813 &0.786 &5350 &0.354 &0.320 &7350 &0.934 &0.890 &8650 &0.935 &0.778 \\
3600 &0.937 &0.506 &4100 &0.107 &0.053 &4600 &0.839 &0.803 &5400 &0.572 &0.521 &7400 &0.980 &0.914 &8700 &0.979 &0.837 \\
3650 &0.805 &0.413 &4150 &0.039 &0.019 &4650 &0.856 &0.813 &5450 &0.772 &0.717 &7450 &0.993 &0.927 &8750 &0.995 &0.876 \\
3700 &0.595 &0.291 &4200 &0.015 &0.007 &4700 &0.876 &0.827 &5500 &0.951 &0.893 &7500 &1.000 &0.936 &8800 &1.000 &0.909 \\
3750 &0.332 &0.155 &4250 &0.000 &0.000 &4750 &0.896 &0.839 &5550 &0.932 &0.902 &7550 &0.936 &0.942 &8850 &0.982 &0.927 \\
3800 &0.113 &0.051 &     &      &      &4800 &0.918 &0.854 &5600 &0.953 &0.906 &7600 &0.709 &0.948 &8900 &0.946 &0.944 \\
3850 &0.015 &0.006 &     &      &      &4850 &0.939 &0.866 &5650 &0.962 &0.926 &7650 &0.716 &0.950 &8950 &0.845 &0.952 \\
3900 &0.000 &0.000 &     &      &      &4900 &0.953 &0.870 &5700 &0.926 &0.885 &7700 &0.917 &0.946 &9000 &0.773 &0.959 \\
     &      &      &     &      &      &4950 &0.971 &0.880 &5750 &0.943 &0.895 &7750 &0.959 &0.934 &9050 &0.764 &0.968 \\ 
6950 &0.000 &0.0002 &     &      &      &5000 &0.981 &0.882 &5800 &0.947 &0.905 &7800 &0.945 &0.920 &9100 &0.719 &0.965 \\
7000 &0.002 &0.0006 &     &      &      &5050 &0.990 &0.884 &5850 &0.937 &0.895 &7850 &0.934 &0.913 &9150 &0.699 &0.973 \\ 
7050 &0.005 &0.0014 &     &      &      &5100 &0.995 &0.883 &5900 &0.946 &0.901 &7900 &0.927 &0.911 &9200 &0.708 &0.974 \\
7100 &0.010 &0.0026 &     &      &      &5150 &1.000 &0.883 &5950 &0.947 &0.905 &7950 &0.923 &0.916 &9250 &0.661 &0.972 \\
7150 &0.014 &0.0038 &     &      &      &5200 &0.997 &0.877 &6000 &0.927 &0.891 &8000 &0.931 &0.934 &9300 &0.493 &0.985 \\
7200 &0.014 &0.0042 &     &      &      &5250 &0.988 &0.865 &6050 &0.919 &0.881 &8050 &0.929 &0.943 &9350 &0.348 &0.980 \\
7250 &0.011 &0.0033 &     &      &      &5300 &0.980 &0.855 &6100 &0.898 &0.867 &8100 &0.902 &0.951 &9400 &0.361 &0.991 \\
7300 &0.008 &0.0022 &     &      &      &5350 &0.960 &0.835 &6150 &0.895 &0.858 &8150 &0.826 &0.938 &9450 &0.350 &0.983 \\
7350 &0.006 &0.0016 &     &      &      &5400 &0.937 &0.811 &6200 &0.926 &0.883 &8200 &0.784 &0.917 &9500 &0.341 &0.979 \\
7400 &0.002 &0.0005 &     &      &      &5450 &0.908 &0.783 &6250 &0.945 &0.903 &8250 &0.733 &0.854 &9550 &0.350 &0.977 \\
7450 &0.000 &0.0002 &     &      &      &5500 &0.874 &0.751 &6300 &0.966 &0.919 &8300 &0.643 &0.751 &9600 &0.368 &0.976 \\
     &      &      &     &      &      &5550 &0.828 &0.710 &6350 &0.974 &0.931 &8350 &0.565 &0.647 &9650 &0.407 &0.986 \\
     &      &      &     &      &      &5600 &0.775 &0.663 &6400 &0.947 &0.904 &8400 &0.576 &0.627 &9700 &0.432 &0.981 \\
     &      &      &     &      &      &5650 &0.719 &0.614 &6450 &0.970 &0.909 &8450 &0.734 &0.772 &9750 &0.424 &0.989 \\
     &      &      &     &      &      &5700 &0.658 &0.561 &6500 &1.000 &0.941 &8500 &0.554 &0.871 &9800 &0.402 &0.990 \\
     &      &      &     &      &      &5750 &0.594 &0.506 &6550 &0.928 &0.884 &8550 &0.123 &0.216 &9850 &0.378 &0.984 \\
     &      &      &     &      &      &5800 &0.524 &0.447 &6600 &0.907 &0.835 &8600 &0.028 &0.047 &9900 &0.355 &0.981 \\
     &      &      &     &      &      &5850 &0.460 &0.392 &6650 &0.986 &0.894 &8650 &0.009 &0.014 &9950 &0.334 &0.980 \\
     &      &      &     &      &      &5900 &0.391 &0.334 &6700 &0.990 &0.917 &8700 &0.004 &0.006 &10000&0.312 &0.981 \\
     &      &      &     &      &      &5950 &0.329 &0.281 &6750 &0.923 &0.853 &8750 &0.002 &0.003 &10050&0.291 &0.982 \\
     &      &      &     &      &      &6000 &0.271 &0.232 &6800 &0.858 &0.793 &8800 &0.000 &0.003 &10100&0.270 &0.981 \\
     &      &      &     &      &      &6050 &0.219 &0.188 &6850 &0.782 &0.740 &     &      &      &10150&0.250 &0.987 \\
     &      &      &     &      &      &6100 &0.173 &0.148 &6900 &0.849 &0.787 &     &      &      &10200&0.225 &0.977 \\
     &      &      &     &      &      &6150 &0.135 &0.116 &6950 &0.631 &0.676 &     &      &      &10250&0.203 &0.975 \\
     &      &      &     &      &      &6200 &0.103 &0.088 &7000 &0.210 &0.249 &     &      &      &10300&0.181 &0.976 \\
     &      &      &     &      &      &6250 &0.077 &0.066 &7050 &0.093 &0.096 &     &      &      &10350&0.159 &0.973 \\
     &      &      &     &      &      &6300 &0.057 &0.049 &7100 &0.071 &0.065 &     &      &      &10400&0.134 &0.967 \\
     &      &      &     &      &      &6350 &0.042 &0.036 &7150 &0.052 &0.056 &     &      &      &10450&0.112 &0.963 \\
     &      &      &     &      &      &6400 &0.031 &0.026 &7200 &0.017 &0.023 &     &      &      &10500&0.091 &0.970 \\
     &      &      &     &      &      &6450 &0.020 &0.017 &7250 &0.006 &0.008 &     &      &      &10600&0.048 &0.981 \\
     &      &      &     &      &      &6500 &0.014 &0.012 &7300 &0.000 &0.004 &     &      &      &10700&0.000 &0.989 \\
     &      &      &     &      &      &6550 &0.012 &0.010 &     &      &      &     &      &	   &     &      &      \\
     &      &      &     &      &      &6600 &0.007 &0.006 &     &      &      &     &      &	   &     &      &      \\
     &      &      &     &      &      &6650 &0.005 &0.004 &     &      &      &     &      &	   &     &      &      \\
     &      &      &     &      &      &6700 &0.000 &0.002 &     &      &      &     &      &	   &     &      &      \\
\enddata
 \end{deluxetable}

{\it Facilities:} \facility{Skymapper ()}
 

\end{document}